# Epidemiology of Objectively Measured Bedtime and Chronotype in the US adolescents and adults: NHANES 2003-2006

Jacek K. Urbanek, Adam Spira, Junrui Di, Andrew Leroux, Ciprian Crainiceanu, Vadim Zipunnikov

## ABSTRACT

**Background:** We propose a method for estimating the timing of in-bed intervals using objective data in a large representative U.S. sample, and quantify the association between these intervals and age, sex, and day of the week.

**Methods:** The study included 11,951 participants six years and older from the National Health and Nutrition Examination Survey (NHANES) 2003-2006, who wore accelerometers to measure physical activity for seven consecutive days. Participants were instructed to remove the device just before the nighttime sleep period and put it back on immediately after. This nighttime period of non-wear was defined in this paper as the objective bedtime (OBT), an objectively estimated record of the in-bed-interval. For each night of the week, we estimated two measures: the duration of the OBT (OBT-D) and, as a measure of the chronotype, the midpoint of the OBT (OBT-M). We estimated day-of-the-week-specific OBT-D and OBT-M using gender-specific population percentile curves. Differences in OBT-M (chronotype) and OBT-D (the amount of time spent in bed) by age and sex were estimated using regression models.

**Results:** The estimates of OBT-M and their differences among age groups were consistent with the estimates of chronotype obtained via self-report in European populations. The average OBT-M varied significantly by age, while OBT-D was less variable with age. In the reference group (females, aged 17-22 years), the average OBT-M across 7 days was 4:19AM (SD = 30 min) and the average OBT-D was 9h 19min (SD=12 min). In the same age group the average OBT-D was 18 minutes shorter for males than for females, while the average OBT-M was not significantly different between males and females. The most pronounced differences were observed between OBT-M of weekday and weekend nights. In the reference group, compared to the average OBT-M of 3:50am on Monday through Thursday nights, there was a 57-minute delay in OBT-M on Friday nights (entering the weekend), a 69-minute delay on Saturday nights (staying in the weekend), and a 23-minute delay on Sunday night (leaving the weekend). For both OBT-M and OBT-D, in most age groups and for most days of the week, there were no statistically significant differences between males and females, except for OBT-D on Wednesdays and Thursdays, with males having 31 (p-value < 0.05) and 45 (p-value < 0.05) minutes shorter OBT-D, respectively.

**Conclusions:** The proposed measures, OBT-D and OBT-M, provide useful information of time in bed and chronotype in NHANES 2003-2006. They identify within-week patterns of bedtime and can be used to study associations between the bedtime and the large number of health outcomes collected in NHANES 2003-2006.

# INTRODUCTION

The importance of sleep and its influence on health is well documented with research interest in sleep and biological rhythms continuing to increase across health research disciplines. Many large epidemiological studies and clinical trials have collected or are in the process of collecting data on sleep and associated health outcomes (Blackwell et al., 2011; Redline et al., 2014; Schrack et al., 2013; Spira et al., 2013). In particular, the timing and duration of sleep have been identified as strong predictors of follow-up comorbidities (Czeisler, 2015; Skeldon, Derks, & Dijk, 2016; Tranah et al., 2011; Wulff, Gatti, Wettstein, & Foster, 2010; Yaffe et al., 2011). Interestingly, both short and long sleep duration may be associated with adverse health outcomes, including metabolic syndrome, obesity, heart disease, and mortality (Aurora, Kim, Crainiceanu, O'Hearn, & Punjabi, 2016; Baron, Reid, Kern, & Zee, 2011; Cappuccio, Cooper, D'Elia, Strazzullo, & Miller, 2011; Grandner, Hale, Moore, & Patel, 2010; Hsieh, Muto, Murase, Tsuji, & Arase, 2011; Wu, Zhai, & Zhang, 2014). The timing of sleep, hormone release and core body temperature are regulated in large part by the circadian clock (Fuller, Gooley, & Saper, 2006; Gangwisch, 2009), and dysregulation of the circadian clock has been linked to obesity, diabetes, mood, sleep disorders and other comorbidities (Gottlieb et al., 2005; Grandin, Alloy, & Abramson, 2006; Lu & Zee, 2006; McClung, 2007).

Many large epidemiological studies collect data on sleep duration and timing using self-report questionnaires (Freedman et al., 2011; Spira et al., 2013). For example, the Munich ChronoType Questionnaire (MCTQ) collects information about when respondents prepare for sleep and go to bed, how long they sleep, when they wake up, and when they get out of bed on work and non-work days (Juda, Vetter, & Roenneberg, 2013b; Roenneberg et al., 2007). Results based on the MCTQ are used to characterize sleep chronotype, a personal preference for timing of sleep. To mitigate the influence of sleep duration, the mid-point of the sleep interval is commonly used as a measure of chronotype (Roenneberg et al., 2007; Wittmann, Dinich, Merrow, & Roenneberg, 2006). Based on the mid-point, two major sleep chronotypes are defined: morning type (go to bed earlier) and evening type (go to bed later).

Chronotype and sleep duration have been studied in different age and disease cohorts (Borisenkov, 2010; Borisenkov, Perminova, & Kosova, 2010; Roenneberg et al., 2007; Wittmann et al., 2006). However, only limited research has been dedicated to objective measures of sleep parameters in nationally representative samples. We propose to objectively estimate the chronotype and time in bed in the National Health and Nutrition Examination Survey (NHANES) 2003-2006. NHANES is a cross-sectional, US nationally representative survey that assessed demographic, dietary and health-related questions to understand differences in health and nutrition across age groups (Troiano et al., 2008). To obtain objective measurements we analyzed nighttime non-wear periods in data from a physical activity monitor. Previous analyses of NHANES data (Healy, Matthews, Dunstan, Winkler, & Owen, 2011; Tucker, Welk, & Beyler, 2011; Tudor-Locke et al., 2011) treated the non-wear data as "missing" because they focused on physical activity rather than sleep. *We propose to revisit these data that were thought not to contain relevant information and treat non-wear periods as implicit measures of objective bedtime period (OBT).* Using this approach for the NHANES 2003-2006 accelerometry data, we computed U.S. nationally representative estimates of chronotype and amount of time spent in bed, by age and sex. We also estimated day-of-the-week-specific duration of the OBT (OBT-D) and the midpoint of the OBT (OBT-M) as well as population percentile curves for each gender. Further, we validated our OBT-M index against results reported for midpoints of sleep in European populations using self-administered questionnaires (Juda, Vetter, & Roenneberg, 2013a; Juda et al., 2013b; Wittmann et al., 2006).

# METHODS

The NHANES is a cross-sectional, nationally representative survey conducted to better understand differences in health and nutrition across age groups in the U.S. Most NHANES data are publicly available on the website of the National Center for Health Statistics (NCHS) (https://www.cdc.gov/hchs/nhanes). All participants or their legally authorized representative provided informed consent, and the NCHS Ethics Review Board approved all survey protocols. NCHS provided sample weights to allow valid population estimates for specifically defined demographic groups. All results reported in this paper account for the NHANES sampling scheme and are representative of the 2003-2006 US population.

*Participants*

NHANES 2003-2006 included a representative sample of the US civilian non-institutionalized population selected with a complex, multistage probability design. Here we use data from participants aged 6 years and older. Participants were excluded from the accelerometry portion of the study if: 1) their waist was too large for the accelerometer belt; 2) they were in a wheelchair; 3) they had recently undergone abdominal surgery; or 4) if they were younger than 6 years of age. In total, 14,631 participants completed the accelerometry portion of the study among which 11,951 participants had at least one valid day. Table A.1 in the Appendix reports the details on the number of participants with valid days across age groups and days of the week.

*Measures*

**Accelerometry.** NHANES participants were provided an Actigraph 7164 uni-axial accelerometer (ActiGraph, Pensacola, FL) to wear on the right hip with an elastic belt. Participants were asked to wear the device for 7 consecutive days during the daytime and to take it off for swimming, bathing and showering. *They were instructed to remove the monitor before getting into bed and put it back on just after getting out of bed.* After completing data collection, participants were instructed to return the device by mail. The activity monitor recorded movement intensity values, expressed in counts per minute (CPM). The NCHS and survey collaborators performed an initial data review to identify outliers and unreasonable values. Additionally, we excluded days with 10 or more hours of non-wear time. We defined non-wear periods as intervals of at least 60 consecutive minutes of zero activity counts, with allowance for 1-2 min of counts between 0 and 100 (Atienza et al., 2011; Troiano et al., 2008). For automated identification of non-wear periods we applied the algorithm included in NHANESACCEL R-package (Van Domelen & Pittard, 2014).

**Sleep questionnaire.** Additionally, for comparison to accelerometer non-wear-based estimates, we used data from the sleep disorders questionnaire used in NHANES 2005-2006 [https://wwwn.cdc.gov/Nchs/Nhanes/2005-2006/SLQ_D.htm]. We used self-reported average sleep duration ("How much sleep do you get (hours)?") and self-reported average sleep onset latency ("How long to fall asleep (minutes)?"). Participants reported average sleep duration between 1 and 11 hours. Sleep duration longer than 11 hours was categorized as 12 hours and longer. Sleep onset latency was reported between 0 to 50 minutes. Values above 50 minutes were categorized as 60 minutes and longer. For this portion of the study, the NHANES protocol focused on the eligible sample of participants 16 years and older. We only use sleep-questionnaire data for participants who completed accelerometry, resulting in 4,357 participants.

**Other measures.** NHANES participants reported their age, sex, and ethnicity at enrollment. Their weight and height have been measured during the examination. We only used data on age and sex for analyses. Detailed characteristic of study participants across age groups are presented in Table 1.

*Analyses*

**Objective Bedtime (OBT).** To estimate OBT, we calculated the duration of the longest non-wear period within a 24-hour window, which ranged from 4 to 14 hours. Non-wear duration was restricted to this range to avoid confusion with short periods that are not related to bedtime and long periods that are likely to be invalid data. Only valid days were considered, and only non-wear periods that were preceded and followed by recorded physical activity. If a non-wear period started on an invalid day, or before the 7-day assessment period, it was treated as invalid. Visual examples of valid and invalid OBT estimated using accelerometry data are displayed in Figure 1.

**Assignment of OBT to a day of the week.** We assigned each OBT to a specific day of the week based on the timing of the OBT. If the OBT ended before noon, it was assigned to the day that preceded the end of that OBT. If the OBT ended in the afternoon, the OBT was assigned to the day on which the period ended. For example, if the OBT started on Monday at 11PM and ended on Tuesday at 7AM, the interval was assigned to Monday night. If the OBT started on Tuesday at 9AM and ended on Tuesday at 1PM, the OBT was assigned to Tuesday night. Thus, the more typical "nighttime" OBT was assigned to the night on which the OBT was initiated, whereas the less typical "daytime" OBT (participants with sleep disorders, shift workers etc.) was assigned to the day the OBT actually began.

**Features of OBT.** We focused on two features of OBT: duration and midpoint. We defined the estimated duration of OBT (OBT-D) as the length of the time interval between the beginning and end of the identified OBT expressed in complete hours followed by minutes. For example OBT-D equal to 9 hours and thirty minutes is denoted as 9:30. We defined the midpoint of OBT (OBT-M) as the midpoint between the beginning and end of the OBT period, expressed as time of day (e.g. 4:30AM). Figure 1 provides a graphical representation of invalid periods, OBT duration, and OBT midpoint. To distinguish features of OBT between different days of the week, we used the lower-index notation. For example, duration of OBT observed on Monday night is denoted by OBT-$D_{Mon}$, while midpoint of OBT observed on Sunday night is denoted by OBT-$M_{Sun}$.

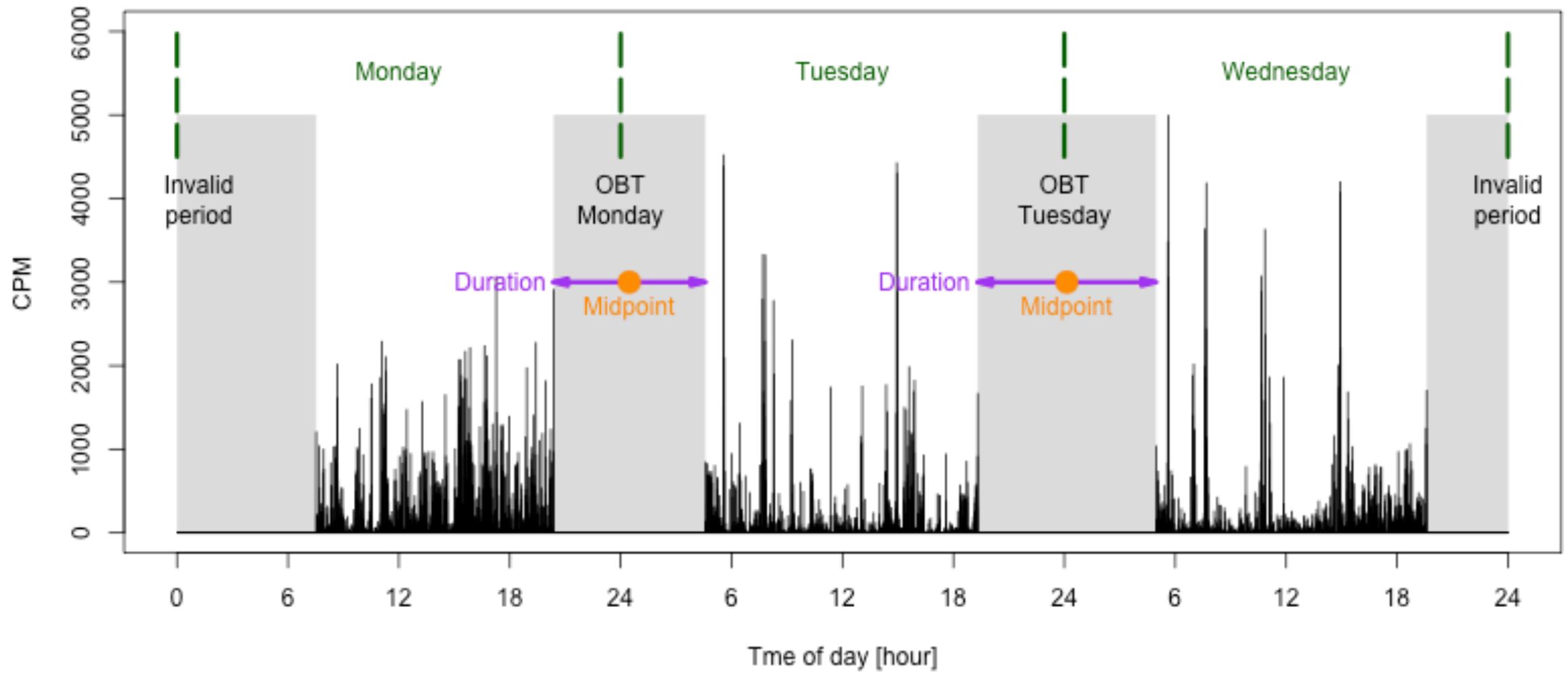

Figure. 1. Graphical representation of OBT. Black vertical lines represent activity counts per minute (CPM), gray areas represent non-wear periods, green dashed lines separate days of the week, purple arrows indicate OBT durations and orange points represent OBT midpoints. The invalid period on the left was considered invalid because no physical activity data were recorded before it. The invalid period on the right was considered invalid because no physical activity data appear after it.

**Population percentile curves.** We estimated the age- and sex-specific 5%, 10%, 25%, 50%, 75%, 90%, 95% percentile curves for the OBT measures in NHANES. Figures in the paper focus on the median (50%) curves, while the other percentile curves are reported in the Appendix. Generalized additive models for location, scale and shape (GAMLSS) (Rigby & Stasinopoulos, 2006; Stasinopoulos & Rigby, 2007) using the Lambda-Mu-Sigma (LMS) method were used to estimate these curves (Cole

& Green, 1992; Flegal & Cole, 2013). LMS transforms data for each age using a Box-Cox transformation with transformation parameters that depend smoothly on age. The transformed data are modeled using a normal or $t$ distribution and quantiles are fitted to the transformed data and mapped back onto the original scale. The LMS model was fitted using survey-weighted penalized likelihood (Cole & Green, 1992). For computational stability, survey weights were normalized to sum to the sample size. All analyses were performed using the "survey" package in R (Lumley, 2011).

**Regression analysis.** We modeled OBT-M and OBT-D as a function of gender, age, and the gender-by-age interaction separately for each day of the week. To capture the heterogeneous patterns observed in Figures 2-5, thirteen age groups, shown in Table 1, were created and the age group indicators were included in all models to represent the age effect. Motivated by the patterns observed in Figures 2-5, indicating a low in OBT-D and a high in OBT-M between 17-22 years of age, the 17-22 year old females were chosen to be the reference group. Seven day-of-the-week-specific models were fitted both for OBT-D and OBT-M as follows:

$$Y_i = g_F + g_M I_{\{Gender_i=Male\}} + \sum_{l=1}^{12} a_l I_{\{Age_i \in Age^l\}} + \sum_{l=1}^{12} a_l I_{\{Age_i \in Age^l\}} I_{\{Gender_i=Male\}},$$

where $Y_i$ denotes OBT-D or OBT-M of the i-th NHANES participant having gender $Gender_i$ and $Age_i$. Thus, $g_F$ and $g_F + g_M$ are the parameters for a female and male, respectively, from the 17-22 years old reference age group. Similarly, $g_F + a_l$ and $g_F + g_M + a_l + ga_l$ are the parameters for a female and male, respectively, from the age group $Age^l$. This approach allows testing for gender and age differences within a specific day of the week and provides estimates of the group-specific mean effects.

## RESULTS

For presentation purposes we provide and interpret only the median (50%) curves of OBT-M and OBT-D over the lifespan; the other percentile population curves are presented in the Appendix (figures A.1 and A.2).

We focus first on comparing OBT-D and self-reported sleep duration in the participants of the 2005-06 NHANES older than 16 years of age, who have both of these measures. The subject-specific OBT-D represents an average across all available days for that subject. The left panel in Figure 2 displays age-specific medians of self-reported sleep durations (dashed lines) for males (blue) and females (red), together with medians of subject-specific OBT-D (solid lines). While visually OBT-D is, on average, 2 hours longer than the self-reported sleep duration, both exhibit similar age-related patterns. The minimum of the median self-reported sleep duration occurs between 40 and 50 years of age for males and between 55 and 65 years of age for females. The minimum of the median OBT-D was between 55 and 65 years of age for both males and females. Possible explanations for this observed difference could be the differential report of sleep onset latency for men and women or different behaviors around sleep time. The right panel in Figure 2 indicates that median self-reported sleep onset remains relatively constant for females over the life span and decreases by about 20% (5 minutes) for males.

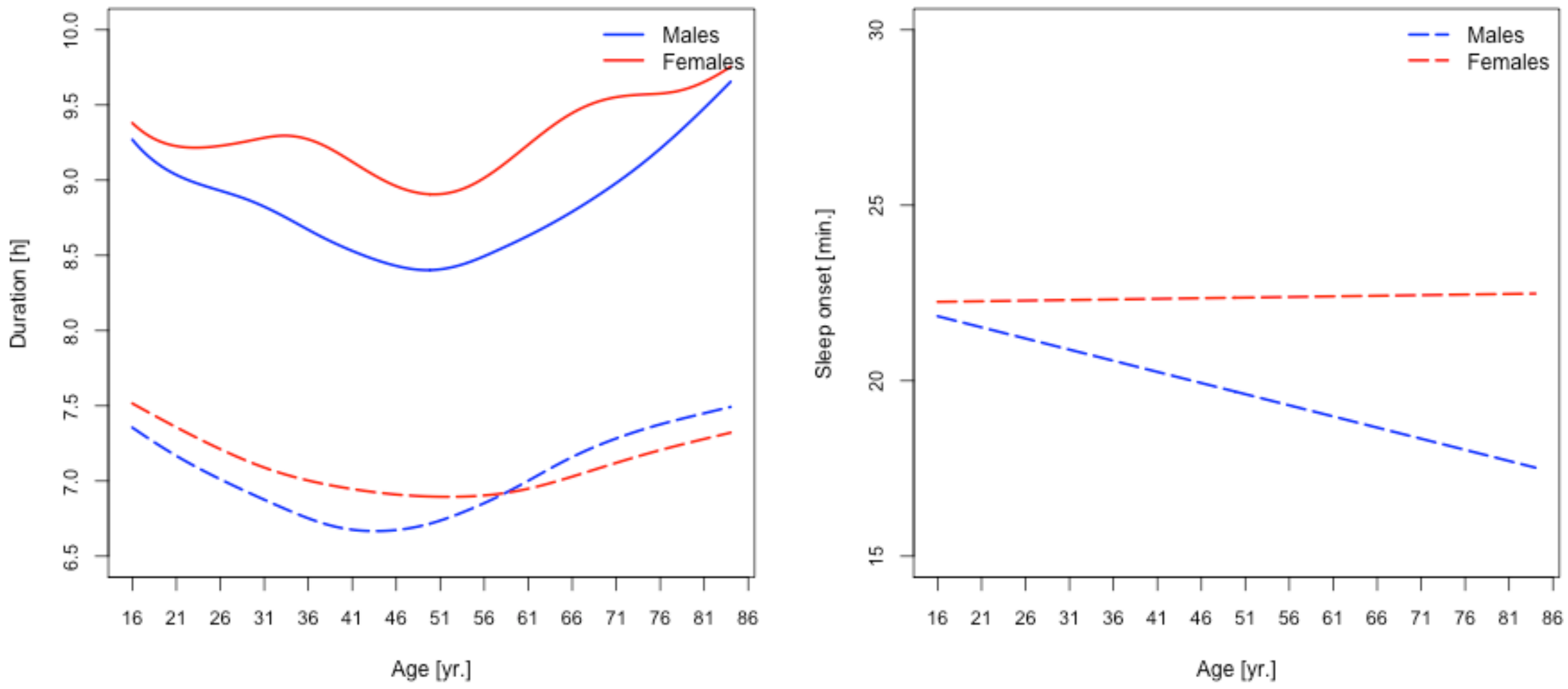

Figure 2. Left panel: median self-reported sleep duration (dashed lines) and estimated duration of OBT-D (solid lines) for males (blue) and females (red). Right panel: median self-reported sleep onset for males (blue) and females (red).

Figure 3 displays the OBT-D medians across the lifespan for each day of the week for males (left panel) and females (right panel). For males, OBT-D decreases as a function of age between the age of 6 and 20 and then exhibits a slight increase as a function of age, though patterns exhibit more heterogeneity depending on the day of the week. Indeed, weekend median OBT-D exhibits a clearer and steadier increase, while weekdays median OBT-D exhibit a wavier pattern with peaks around 30 years of age followed by a slow decrease to around 50 years of age. After 50 years of age the weekdays median OBT-D increases slowly with age to around 9.5 hours. The dissimilarity between weekend and weekdays OBT-D is different for females. Indeed, for females the patterns of median OBT-D are more consistent across the lifespan. Moreover, the median OBT for Sunday (OBT-$D_{Sun}$) has a clearly identifiable minimum around 25 years of age for males, while for females there are two such minima, one around age 16 and one around age 46.

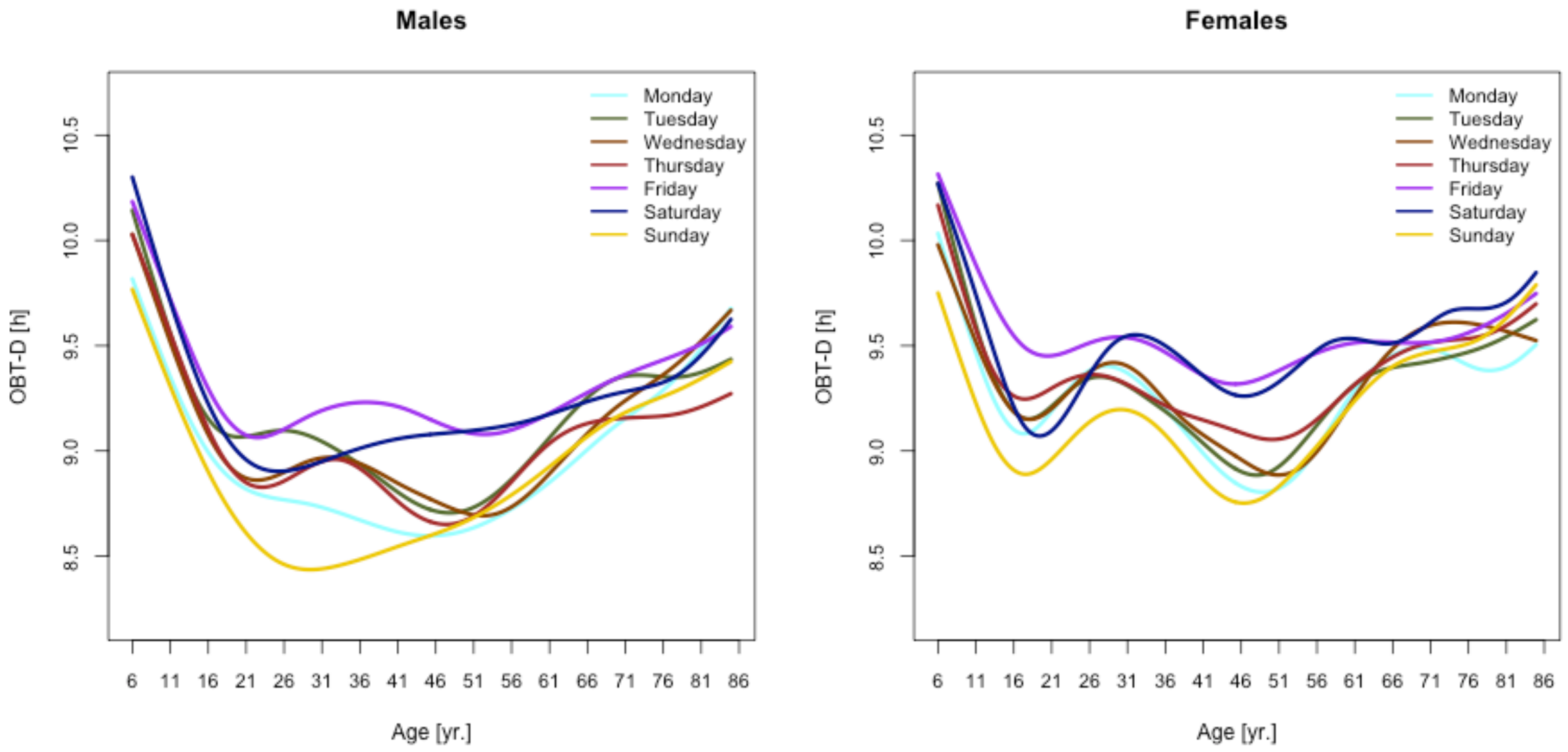

Figure. 3. OBT-D estimated for each day of the week for males (left) and females (right).

The results of the estimated regression model (1) for OBT-D are shown in Table 2. Table 3 shows the fitted values obtained from the estimated regression models for OBT-D for both genders and all age groups. For Wednesdays, the average OBT-D of a female in the reference 17-22 years old age group is 9 hours and 24 four minutes, which is 31 minutes longer than the OBT-D of a male from the same group. Compared to the reference group, the 6-10 years old age group has an average OBT-D across all days of the week betwen 40-80 minutes longer. Only Fridays and Saturdays nights have 36 and 51 minutes longer OBT-D in the 11-16 years old group compared to the reference group. Across the life span, Saturday OBT-D seems to be the longest and, compared to the reference group, the difference increases by 17 to 70 minutes, depending on the age group. Gender differences are stronger on Wednesdays and Thursdays (all age groups) and Saturdays (23-40 years old and 71-76 years old age groups). The results are mostly consistent with the visual intuition provided by the median plots in Figure 3.

Figure 4 displays the medians of OBT-M across the lifespan for weekdays (left panel) and weekends (right panel) for males (blue) and females (red). Estimates for weekends and weekdays are shown separately to compare the OBT-$M_{Weekend}$ and the estimates of self-reported sleep-midpoints in the work-free days in European populations provided in Figure 1.D in Ronnenberg et al. (Roenneberg et al., 2004). Strikingly, the left panel in Figure 4 is almost identical to the Figure 1.D in Ronnenberg et al. (Roenneberg et al., 2004) for the work-free-days. For both weekdays and weekends, the maximum of median OBT-M value is attained around 20 years of age. For weekdays the maximum is around 4AM for both genders, whereas for weekends is around 5:20AM for males and 5AM for females. During weekdays, OBT-M has a slight upward trend after age 35 for both genders, while during weekends OBT-M decreases roughly linearly from around 3:45AM at age 35 to around 3AM at age 85.

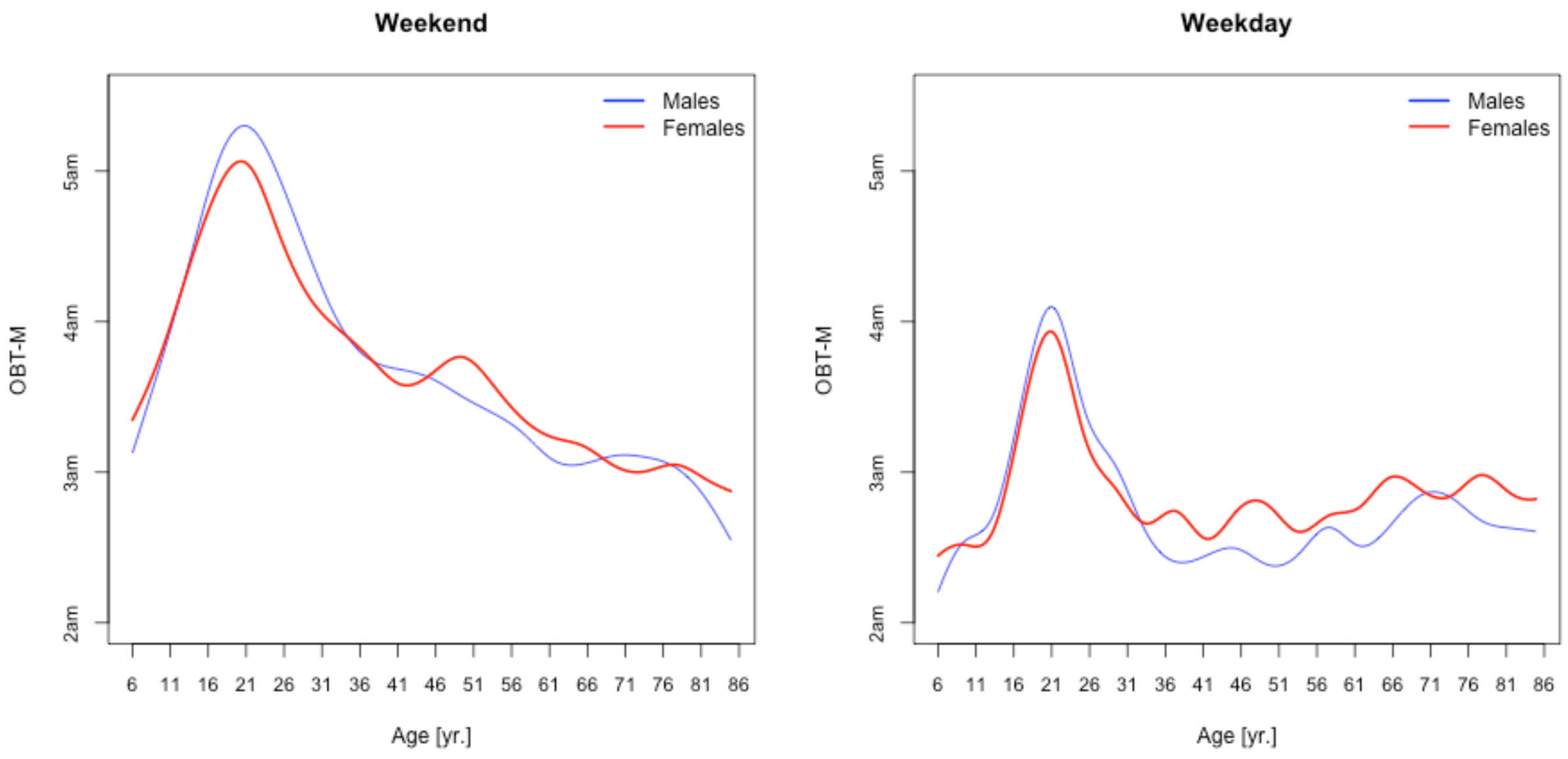

Figure. 4. OBT-M estimated for weekdays (left panel) and weekends (right panel) for males (blue) and females (red).

Figure 5 displays medians of OBT-M estimated across the lifespan for each day of the week for males (left) and females (right). For weekend days the median OBT-M peaks around 20 years of age for both males and females. For weekdays, median OBT-M also peaks around 20 years of age and then starts to decrease to its minimum around 2:30AM around 40 years of age. The median OBT-M stabilizes after 40 years of age for males but increases for females to around 2:45AM at 85 years of age. The maximum value of the median $OBT\text{-}M_{Sat}$ is around 5AM for males and 4:45AM for females. The maximum value of the median $OBT\text{-}M_{Fri}$ is around 4:30AM for males and 4:15AM for females, about 30 minutes less than the values for Saturday. For males around 20 years of age $OBT\text{-}M_{Sun}$ is about 30 minutes later than other weekdays. This difference does not seem to exist for females. The medians OBT-M across weekdays are quite consistent with each other indicating a level of homogeneity of the sleep chronotype across weekdays across the life span with a peak value hovering around 3:30AM for individuals who are around 20 years of age.

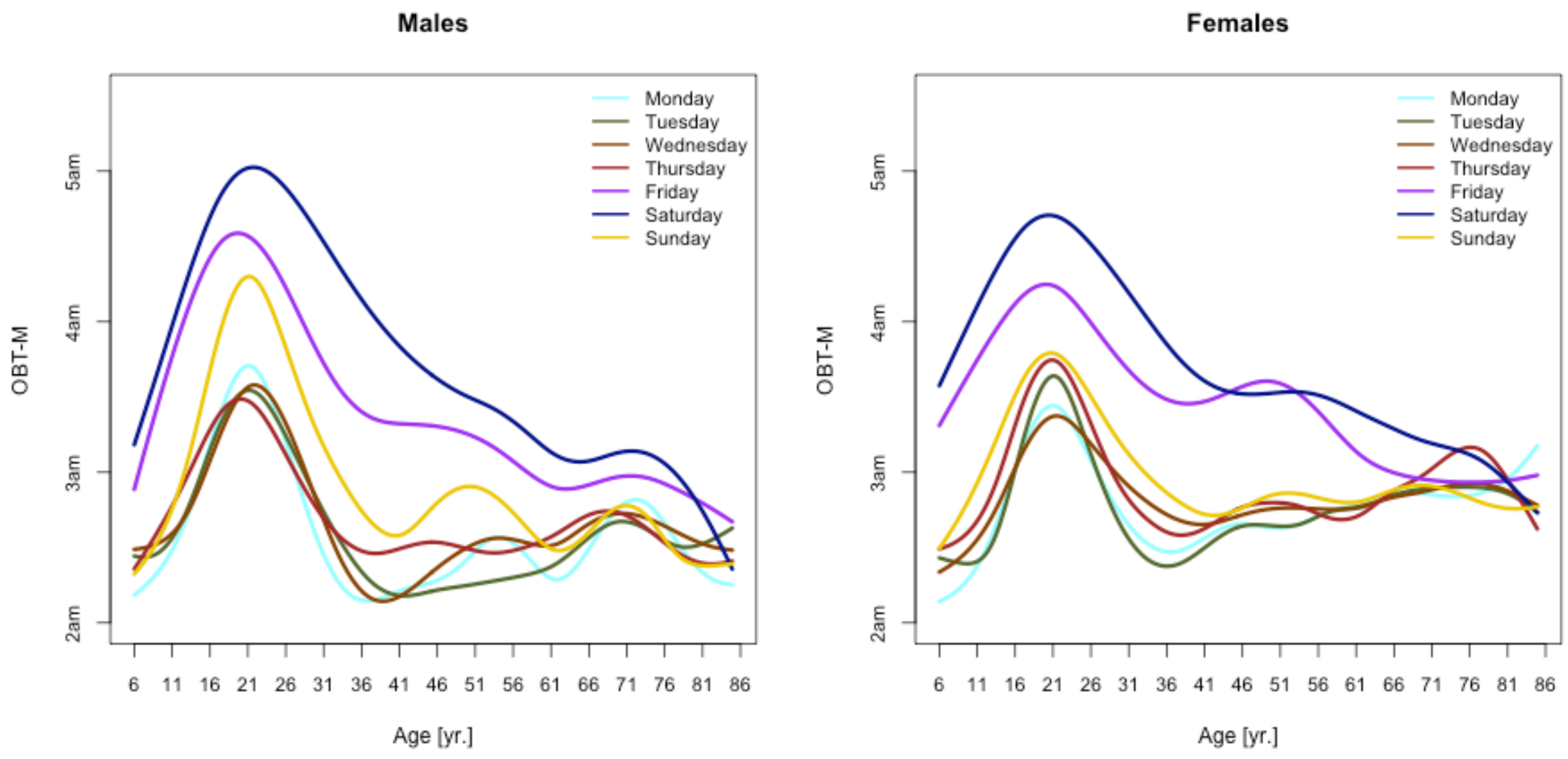

Figure. 5. OBT-M estimated for each day of the week for males (left) and females (right).

Table 4 shows the estimated coefficients for OBT-M using the regression model (1). Table 5 shows the fitted values obtained from the estimated regression models for OBT-M for both genders and all age groups. The main results statistically confirm the patterns observed in Figure 5 over the entire life span. Only on Saturdays, the reference group is not statistically different from the 11-16 years old and 23-29 years old age groups. Males have approximately 40-50 minutes larger OBT-M than females on Sundays (47-52 years old and 65-84 years old age groups) and on Tuesdays (47-52 years old age group).

# DISCUSSION

*In the present study, we developed a novel use for the 2003-2006 NHANES accelerometry data, by interpreting the nighttime non-wear of the accelerometer as an objective measure of the in-bed period.* This approach uses the NHANES accelerometry data that so far has been used exclusively to explore physical activity.

Based on non-wear data we have introduced two new measures of the in-bed period: duration (OBT-D) and midpoint (OBT-M). We showed that these objective non-wear periods could be used in addition to self-reported sleep characteristics when they are available or as a proxy when they are not. Thus, OBT provides objective and quantifiable indirect information about the timing (OBT-M) and duration of the in-bed interval (OBT-D). The weekend OBT-M is strikingly similar to the midpoints of self-reported sleep on work-free days reported by Ronneberg et al., 2004. This suggests that OBT-M is a robust estimator of the midpoint of sleep.

We have also quantified and illustrated the differences in OBT-D and OBT-M over age groups covering most of the lifespan. The differences in OBT-M across the lifespan are statistically significant for all age groups for each day of the week. The differences in OBT-D across age groups are relatively small within adulthood, but are relatively large between children and adults. Our approach allows the objective analysis of OBT periods for each day of the week, which provides an in-depth analysis of within-week chronobiology. This allows characterizing and quantifying the dynamics of the chronobiology of sleep as a function of entering, staying in, and leaving the weekend, relative to the rest of week. These results complement the literature on sleep and bedtime that typically use coarser categories such as week-end/week-day or work/work-free days (Roenneberg et al., 2004; Zavada, Gordijn, Beersma, Daan, & Roenneberg, 2005). Results also indicate that for some age groups, OBT on different days of the week may differ between males and females. For example, OBT-$M_{Sun}$ is statistically significantly different between men and women in the 65 to 84 years old age group with an observed average difference of between 10 to 15 minutes. This suggests that sex-specific OBT-M may be an important factor in aging and health research.

An important limitation of OBT is that it does not contain the activity levels and fragmentation of rest/wake during time-in-bed because the device is simply off during this period. Therefore, the OBT features should be interpreted as measures of bedtime and not of sleep. Another limitation is that not all subjects had seven days of data, which could be due to lack of compliance or equipment failure. Thus, OBT measurements are available for valid days/nights only. Further, we followed the accepted NHANES protocol for estimation of sleep durations by limiting the maximum duration of OBT to 14 hours. This may have resulted in the exclusion of participants with hypersomnia. However, the prevalence of such individuals in the population is expected to be very low (Coleman et al., 1982; Dauvilliers & Buguet, 2005).

Despite these limitations, the present study has several strengths. First, we analyzed data from a very large U.S.-representative sample (11,951 subjects). Another important contribution of our approach is that the OBT features can now be used to link sleep/wake schedules or chronotypes to the multiple NHANES health and mortality outcomes. Moreover, the relative predictive power of these proxies of chronobiology can be used to compare them to other established biomarkers. Future work may focus on studying the association between the proposed OBT measures and health and mortality outcomes in NHANES as well as on estimating systematic within-week variability in the OBT measures including social jet lag.

## Declaration of Interest statement

This work is supported by National Institutes of Health grants: NIH/NHLBI R01HL123407, NIH/NIA R01AG049872-01, NIH/NIA R01AG050507. Dr. Spira is supported by AG049872, AG050507, AG050745, and AG052445 from the National Institute on Aging.

Dr. Spira agreed to serve as a consultant to Awarables, Inc. in support of an NIH grant.

**Table 1.** Survey-weighted demographic characteristics of NHANES participants across 13 age groups.

| | 6-10 y.o. | 11-16 y.o. | 17-22 y.o. | 23-28 y.o. | 29-34 y.o. | 35-40 y.o. | 41-46 y.o. | 47-52 y.o. | 53-58 y.o. | 59-64 y.o. | 65-70 y.o. | 71-76 y.o. | 77-84 y.o. | Overall |
|---|---|---|---|---|---|---|---|---|---|---|---|---|---|---|
| Sex: Male (%) | 51.6 | 51.9 | 50.9 | 49.4 | 49.1 | 48.4 | 48.6 | 48.5 | 49.8 | 45.3 | 46.5 | 44.1 | 42.9 | 48.8 |
| Race (%): | | | | | | | | | | | | | | |
| Mexican American | 13.7 | 13.1 | 11.2 | 13.4 | 13.6 | 10.1 | 8 | 6 | 4.2 | 4.6 | 4 | 3.3 | 2.3 | 8.9 |
| Non-Hispanic - Black | 13.6 | 15.5 | 14.4 | 10.8 | 12.5 | 10.1 | 11.1 | 11.3 | 10.1 | 9.1 | 7.6 | 6.8 | 6.7 | 11.2 |
| Non-Hispanic - White | 59.3 | 61.8 | 64 | 64.7 | 63.6 | 66.7 | 73.8 | 75.2 | 76.5 | 80 | 80.6 | 85.8 | 88.5 | 70.6 |
| Other | 4.3 | 4 | 4.2 | 5.6 | 3.9 | 5.5 | 2.9 | 2.5 | 4.3 | 0.7 | 3.6 | 1.2 | 0.8 | 3.6 |
| Missing | 9.1 | 5.7 | 6.2 | 5.4 | 6.5 | 7.6 | 4.3 | 4.9 | 4.8 | 5.6 | 4.2 | 2.9 | 1.7 | 5.6 |
| Weight[kg]: mean (sd) | 32.25 (10.45) | 59.03 (17.74) | 73.89 (21.09) | 77.73 (20.22) | 81.11 (20.90) | 82.45 (21.77) | 85.70 (22.49) | 82.45 (18.93) | 84.25 (20.11) | 83.27 (20.09) | 81.74 (18.93) | 78.11 (17.83) | 72.69 (14.20) | 75.23 (24.21) |
| BMI: mean (sd) | 17.99 (3.60) | 22.31 (5.34) | 25.50 (6.46) | 26.60 (6.17) | 28.02 (6.36) | 28.55 (6.74) | 29.57 (7.18) | 28.79 (6.33) | 29.32 (6.31) | 29.11 (6.32) | 29.13 (5.79) | 28.23 (5.68) | 26.95 (4.46) | 26.91 (6.94) |

**Table 2**. Estimated regression coefficients (hours:minutes) for the OBT-D. Shaded cells mark statistically significant results ($\alpha \leq 0.05$).

| Females | $a^D_{6-10}$ | $a^D_{11-16}$ | $g^D_F$ | $a^D_{23-28}$ | $a^D_{28-34}$ | $a^D_{35-40}$ | $a^D_{41-46}$ | $a^D_{47-52}$ | $a^D_{53-58}$ | $a^D_{59-64}$ | $a^D_{65-70}$ | $a^D_{71-76}$ | $a^D_{77-84}$ |
|---|---|---|---|---|---|---|---|---|---|---|---|---|---|
| **Sun.** | 0:43 | -0:05 | 9:3 | 0:02 | 0:08 | -0:04 | -0:5 | -0:11 | -0:15 | 0:23 | 0:20 | 0:26 | 0:31 |
| **Mon.** | 0:54 | 0:05 | 9:13 | 0:08 | 0:08 | -0:01 | -0:13 | -0:22 | -0:18 | 0:11 | 0:20 | 0:19 | 0:05 |
| **Tue.** | 0:46 | 0:00 | 9:18 | -0:8 | -0:14 | -0:09 | -0:16 | -0:24 | -0:13 | 0:03 | 0:08 | 0:13 | 0:14 |
| **Wed.** | 0:34 | -0:01 | 9:25 | -0:07 | 0:00 | -0:14 | -0:21 | -0:35 | -0:38 | -0:19 | 0:04 | 0:20 | 0:02 |
| **Thur.** | 0:45 | -0:09 | 9:28 | -0:30 | -0:22 | -0:15 | -0:13 | -0:22 | -0:21 | -0:08 | 0:01 | 0:05 | -0:05 |
| **Fri.** | 1:12 | 0:32 | 9:08 | 0:24 | 0:25 | 0:21 | 0:08 | 0:31 | 0:29 | 0:17 | 0:17 | 0:17 | 0:20 |
| **Sat.** | 1:33 | 1:03 | 8:52 | 0:42 | 0:44 | 0:53 | 0:21 | 0:39 | 0:44 | 0:51 | 0:16 | 1:09 | 0:50 |
| **Males** | $ga^D_{6-10}$ | $ga^D_{11-16}$ | $g^D_M$ | $ga^D_{23-28}$ | $ga^D_{28-34}$ | $ga^D_{35-40}$ | $ga^D_{41-46}$ | $ga^D_{47-52}$ | $ga^D_{53-58}$ | $ga^D_{59-64}$ | $ga^D_{65-70}$ | $ga^D_{71-76}$ | $ga^D_{77-84}$ |
| **Sun.** | 0:04 | 0:30 | -0:10 | -0:21 | -0:21 | -0:15 | -0:04 | -0:01 | 0:04 | -0:24 | -0:07 | -0:04 | -0:10 |
| **Mon.** | 0:11 | 0:20 | -0:21 | -0:9 | -0:18 | -0:11 | 0:06 | -0:02 | 0:14 | -0:15 | -0:15 | 0:04 | 0:21 |
| **Tue.** | 0:06 | 0:10 | -0:09 | 0:23 | 0:04 | -0:05 | -0:20 | -0:12 | -0:13 | -0:18 | 0:06 | -0:05 | -0:06 |
| **Wed.** | 0:32 | 0:28 | -0:32 | 0:11 | 0:13 | 0:16 | 0:08 | 0:16 | 0:23 | 0:29 | 0:05 | -0:03 | 0:38 |
| **Thur.** | 0:32 | 1:01 | -0:44 | 0:48 | 0:52 | 0:23 | 0:04 | 0:06 | 0:21 | 0:34 | 0:17 | 0:25 | 0:26 |
| **Fri.** | 0:00 | 0:16 | -0:13 | -0:02 | -0:12 | -0:02 | -0:02 | -0:20 | -0:20 | 0:08 | -0:05 | 0:14 | 0:15 |
| **Sat.** | -0:25 | -0:37 | 0:18 | -0:50 | -0:56 | -1:09 | -0:25 | -0:48 | -0:55 | -0:47 | -0:10 | -1:04 | -0:50 |

**Table 3**. Fitted values of OBT-D from the regression models (hours:minutes) with standard errors in the brackets (in minutes) for females (upper row) and males (bottom row).

| Females<br>Males | 6-10 y.o. | 11-16 y.o. | 17-22 y.o. | 23-28 y.o. | 29-34 y.o. | 35-40 y.o. | 41-46 y.o. | 47-52 y.o. | 53-58 y.o. | 59-64 y.o. | 65-70 y.o. | 71-76 y.o. | 77-84 y.o. |
|---|---|---|---|---|---|---|---|---|---|---|---|---|---|
| **Sun.** | 9:45(08)<br>9:40(09) | 8:58(08)<br>9:19(07) | 9:03(10)<br>8:53(11) | 9:05(12)<br>8:35(12) | 9:11(11)<br>8:40(11) | 8:58(09)<br>8:34(09) | 8:58(11)<br>8:45(10) | 8:52(10)<br>8:41(11) | 8:48(10)<br>8:42(10) | 9:26(09)<br>8:52(10) | 9:22(09)<br>9:05(08) | 9:29(09)<br>9:15(09) | 9:34(09)<br>9:15(09) |
| **Mon.** | 10:07(06)<br>9:57(06) | 9:18(07)<br>9:17(06) | 9:13(09)<br>8:52(09) | 9:21(09)<br>8:51(11) | 9:20(10)<br>8:42(09) | 9:12(08)<br>8:40(09) | 8:60(09)<br>8:45(10) | 8:51(08)<br>8:28(10) | 8:55(09)<br>8:48(09) | 9:23(09)<br>8:47(10) | 9:33(09)<br>8:57(10) | 9:32(10)<br>9:15(08) | 9:18(09)<br>9:18(09) |
| **Tue.** | 10:04(06)<br>10:00(07) | 9:18(06)<br>9:20(06) | 9:18(09)<br>9:09(09) | 9:10(08)<br>9:24(10) | 9:04(09)<br>8:59(10) | 9:09(09)<br>8:54(08) | 9:02(08)<br>8:33(09) | 8:54(08)<br>8:33(09) | 9:05(10)<br>8:43(09) | 9:21(10)<br>8:54(09) | 9:26(09)<br>9:24(09) | 9:31(08)<br>9:16(09) | 9:32(09)<br>9:17(08) |
| **Wed.** | 9:59(06)<br>9:59(05) | 9:24(05)<br>9:20(05) | 9:25(10)<br>8:53(10) | 9:18(09)<br>8:57(10) | 9:25(09)<br>9:06(09) | 9:11(10)<br>8:55(08) | 9:04(08)<br>8:40(10) | 8:50(08)<br>8:34(09) | 8:47(09)<br>8:38(09) | 9:06(08)<br>9:02(10) | 9:29(09)<br>9:03(09) | 9:45(09)<br>9:10(10) | 9:27(11)<br>9:33(10) |
| **Thur.** | 10:13(07)<br>10:01(06) | 9:19(07)<br>9:35(06) | 9:28(09)<br>8:44(10) | 8:58(10)<br>9:02(11) | 9:06(09)<br>9:14(11) | 9:14(10)<br>8:52(10) | 9:15(09)<br>8:35(10) | 9:06(09)<br>8:28(08) | 9:07(09)<br>8:44(10) | 9:20(09)<br>9:10(11) | 9:30(10)<br>9:03(09) | 9:33(09)<br>9:14(09) | 9:24(10)<br>9:05(09) |
| **Fri.** | 10:19(09)<br>10:06(09) | 9:40(09)<br>9:43(08) | 9:08(11)<br>8:54(11) | 9:32(12)<br>9:17(12) | 9:33(11)<br>9:07(12) | 9:29(11)<br>9:13(11) | 9:15(12)<br>9:00(11) | 9:38(09)<br>9:05(10) | 9:36(09)<br>9:04(11) | 9:25(10)<br>9:20(11) | 9:24(10)<br>9:06(11) | 9:25(10)<br>9:26(10) | 9:28(10)<br>9:29(10) |
| **Sat.** | 10:25(10)<br>10:18(10) | 9:55(07)<br>9:36(09) | 8:52(12)<br>9:10(11) | 9:34(12)<br>9:02(17) | 9:36(13)<br>8:58(14) | 9:45(14)<br>8:54(13) | 9:13(13)<br>9:06(14) | 9:31(11)<br>9:01(11) | 9:36(12)<br>8:59(14) | 9:44(11)<br>9:14(12) | 9:08(09)<br>9:16(10) | 10:01(09)<br>9:15(08) | 9:42(10)<br>9:11(10) |

**Table 4**. Estimated regression coefficients (hours:minutes AM) for the OBT-M. Shaded cells mark statistically significant results ($\alpha \leq 0.05$).

| Females | $a^M_{6-10}$ | $a^M_{11-16}$ | $g^M_F$ | $a^M_{23-28}$ | $a^M_{28-34}$ | $a^M_{35-40}$ | $a^M_{41-46}$ | $a^M_{47-52}$ | $a^M_{53-58}$ | $a^M_{59-64}$ | $a^M_{65-70}$ | $a^M_{71-76}$ | $a^M_{77-84}$ |
|---|---|---|---|---|---|---|---|---|---|---|---|---|---|
| **Sun.** | -1:35 | -1:07 | 4:16 | -1:00 | -0:56 | -1:18 | -1:27 | -1:08 | -1:29 | -1:44 | -1:08 | -1:21 | -1:22 |
| **Mon.** | -1:21 | -1:05 | 3:44 | -0:35 | -1:06 | -1:22 | -0:56 | -0:57 | -0:57 | -0:49 | -0:48 | -0:54 | -0:42 |
| **Tue.** | -1:12 | -1:10 | 3:44 | -0:46 | -1:04 | -1:09 | -1:11 | -0:58 | -0:53 | -0:49 | -0:51 | -0:53 | -0:47 |
| **Wed.** | -1:27 | -1:12 | 3:52 | -0:45 | -0:59 | -1:02 | -1:18 | -1:06 | -1:06 | -0:56 | -0:58 | -0:57 | -0:57 |
| **Thur.** | -1:24 | -0:52 | 3:58 | -0:43 | -1:22 | -1:15 | -1:19 | -1:04 | -1:12 | -1:18 | -1:07 | -0:53 | -0:55 |
| **Fri.** | -1:23 | -0:48 | 4:45 | -0:51 | -0:57 | -1:22 | -1:04 | -0:56 | -1:13 | -1:37 | -1:39 | -1:46 | -1:25 |
| **Sat.** | -1:24 | -0:26 | 5:00 | -0:27 | -0:39 | -1:19 | -1:26 | -1:14 | -1:11 | -1:45 | -1:48 | -1:56 | -1:58 |

| Males | $ga_{6-10}^{M}$ | $ga_{11-16}^{M}$ | $g_{M}^{M}$ | $ga_{23-28}^{M}$ | $ga_{28-34}^{M}$ | $ga_{35-40}^{M}$ | $ga_{41-46}^{M}$ | $ga_{47-52}^{M}$ | $ga_{53-58}^{M}$ | $ga_{59-64}^{M}$ | $ga_{65-70}^{M}$ | $ga_{71-76}^{M}$ | $ga_{77-84}^{M}$ |
|---|---|---|---|---|---|---|---|---|---|---|---|---|---|
| **Sun.** | -0:21 | -0:24 | 0:29 | -0:16 | -0:44 | -0:43 | -0:24 | -0:43 | -0:21 | -0:34 | -0:36 | -0:41 | -1:07 |
| **Mon.** | -0:02 | 0:08 | 0:07 | -0:28 | -0:22 | -0:10 | -0:27 | -0:34 | -0:14 | -0:48 | -0:15 | -0:12 | -0:51 |
| **Tue.** | 0:01 | 0:10 | 0:04 | 0:15 | 0:06 | -0:14 | -0:16 | -0:41 | -0:32 | -0:31 | -0:04 | -0:16 | -0:41 |
| **Wed.** | 0:01 | -0:03 | 0:08 | 0:12 | -0:34 | -0:34 | -0:13 | -0:31 | -0:21 | -0:38` | 0:03 | -0:18 | -0:34 |
| **Thur.** | 0:13 | -0:01 | -0:04 | -0:01 | 0:22 | -0:19 | 0:17 | -0:11 | -0:16 | -0:02 | 0:11 | -0:16 | -0:40 |
| **Fri.** | 0:09 | 0:15 | -0:05 | 0:40 | -0:20 | 0:14 | -0:12 | -0:28 | -0:15 | -0:20 | 0:19 | 0:16 | -0:37 |
| **Sat.** | 0:07 | 0:05 | -0:13 | 0:24 | 0:35 | 0:43 | 0:22 | -0:28 | -0:00 | -0:17 | 0:15 | 0:12 | -0:15 |

**Table 5**. Fitted values of OBT-M from the regression models (hours:minutes AM) with standard errors in the brackets (in minutes) for females (upper row) and males (bottom row).

| Females<br>Males | 6-10 y.o. | 11-16 y.o. | 17-22 y.o. | 23-28 y.o. | 29-34 y.o. | 35-40 y.o. | 41-46 y.o. | 47-52 y.o. | 53-58 y.o. | 59-64 y.o. | 65-70 y.o. | 71-76 y.o. | 77-84 y.o. |
|---|---|---|---|---|---|---|---|---|---|---|---|---|---|
| **Sun.** | 2:41AM(07)<br>2:49AM(07) | 3:09AM(10)<br>3:14AM(08) | 4:16AM(12)<br>4:45AM(12) | 3:16AM(14)<br>3:29AM(20) | 3:20AM(11)<br>3:05AM(15) | 2:58AM(08)<br>2:43AM(18) | 2:49AM(11)<br>2:54AM(14) | 3:07AM(09)<br>2:53AM(12) | 2:47AM(09)<br>2:54AM(13) | 2:32AM(11)<br>2:26AM(10) | 3:07AM(06)<br>3:01AM(07) | 2:55AM(07)<br>2:42AM(08) | 2:53AM(07)<br>2:16AM(11) |
| **Mon.** | 2:24AM(04)<br>2:28AM(07) | 2:39AM(08)<br>2:54AM(06) | 3:44AM(09)<br>3:51AM(16) | 3:09AM(12)<br>2:48AM(18) | 2:38AM(13)<br>2:23AM(14) | 2:22AM(11)<br>2:19AM(13) | 2:48AM(09)<br>2:27AM(12) | 2:47AM(09)<br>2:19AM(13) | 2:47AM(09)<br>2:40AM(12) | 2:55AM(09)<br>2:14AM(09) | 2:56AM(07)<br>2:47AM(08) | 2:51AM(09)<br>2:45AM(08) | 3:02AM(07)<br>2:18AM(11) |
| **Tue.** | 2:32AM(04)<br>2:37AM(05) | 2:34AM(07)<br>2:48AM(06) | 3:44AM(12)<br>3:48AM(15) | 2:58AM(15)<br>3:17AM(14) | 2:40AM(07)<br>2:50AM(11) | 2:35AM(08)<br>2:25AM(11) | 2:34AM(08)<br>2:21AM(12) | 2:46AM(09)<br>2:09AM(11) | 2:51AM(09)<br>2:23AM(11) | 2:55AM(11)<br>2:28AM(09) | 2:53AM(06)<br>2:52AM(08) | 2:51AM(08)<br>2:39AM(12) | 2:58AM(06)<br>2:20AM(12) |
| **Wed.** | 2:25AM(07)<br>2:34AM(05) | 2:40AM(07)<br>2:45AM(07) | 3:52AM(10)<br>4:00AM(14) | 3:07AM(13)<br>3:27AM(12) | 2:53AM(08)<br>2:27AM(15) | 2:50AM(10)<br>2:24AM(09) | 2:34AM(07)<br>2:29AM(10) | 2:47AM(09)<br>2:24AM(11) | 2:46AM(10)<br>2:33AM(12) | 2:56AM(09)<br>2:27AM(09) | 2:54AM(07)<br>3:05AM(07) | 2:55AM(07)<br>2:45AM(10) | 2:55AM(08)<br>2:29AM(11) |
| **Thur.** | 2:34AM(06)<br>2:43AM(05) | 3:06AM(08)<br>3:01AM(07) | 3:58AM(12)<br>3:55AM(15) | 3:15AM(12)<br>3:10AM(12) | 2:36AM(12)<br>2:54AM(12) | 2:43AM(09)<br>2:21AM(12) | 2:39AM(12)<br>2:52AM(10) | 2:54AM(07)<br>2:40AM(11) | 2:47AM(07)<br>2:27AM(14) | 2:40AM(08)<br>2:35AM(09) | 2:51AM(10)<br>2:58AM(07) | 3:05AM(08)<br>2:45AM(11) | 3:03AM(09)<br>2:19AM(12) |
| **Fri.** | 3:22AM(12)<br>3:26AM(11) | 3:57AM(09)<br>4:07AM(09) | 4:45AM(12)<br>4:40AM(17) | 3:55AM(15)<br>4:29AM(16) | 3:48AM(10)<br>3:23AM(20) | 3:24AM(11)<br>3:32AM(15) | 3:41AM(13)<br>3:24AM(12) | 3:49AM(09)<br>3:16AM(12) | 3:33AM(08)<br>3:12AM(14) | 3:08AM(09)<br>2:43AM(11) | 3:06AM(10)<br>3:19AM(11) | 2:59AM(09)<br>3:10AM(09) | 3:20AM(08)<br>2:38AM(12) |
| **Sat.** | 3:36AM(07)<br>3:31AM(09) | 4:34AM(07)<br>4:27AM(09) | 5:00AM(16)<br>4:48AM(25) | 4:34AM(14)<br>4:45AM(18) | 4:21AM(11)<br>4:43AM(19) | 3:41AM(12)<br>4:11AM(19) | 3:35AM(23)<br>3:44AM(20) | 3:46AM(12)<br>3:06AM(16) | 3:49AM(08)<br>3:36AM(15) | 3:15AM(08)<br>2:45AM(12) | 3:12AM(08)<br>3:15AM(07) | 3:04AM(07)<br>3:03AM(10) | 3:02AM(08)<br>2:34AM(13) |

# Appendix

Table.A.1 Number of participants per age group and day of the week. Values in brackets represent percentage of females. Last column represents number of subjects per age group.

| | 6-10 y.o. | 11-16 y.o. | 17-22 y.o. | 23-28 y.o. | 29-34 y.o. | 35-40 y.o. | 41-46 y.o. | 47-52 y.o. | 53-58 y.o. | 59-64 y.o. | 65-70 y.o. | 71-76 y.o. | 77-84 y.o. | |
|---|---|---|---|---|---|---|---|---|---|---|---|---|---|---|
| **Sun.** | 568 (53%) | 922 (49%) | 592 (51%) | 307 (57%) | 314 (55%) | 318 (49%) | 335 (51%) | 331 (48%) | 284 (50%) | 404 (54%) | 421 (46%) | 352 (46%) | 363 (48%) | |
| **Mon.** | 767 (50%) | 1242 (49%) | 740 (49%) | 407 (55%) | 432 (53%) | 450 (48%) | 450 (50%) | 431 (51%) | 353 (50%) | 454 (53%) | 428 (50%) | 359 (47%) | 347 (47%) | |
| **Tue.** | 805 (51%) | 1353 (49%) | 768 (49%) | 449 (59%) | 460 (52%) | 470 (48%) | 508 (50%) | 475 (51%) | 385 (50%) | 493 (51%) | 465 (49%) | 376 (47%) | 374 (50%) | |
| **Wed.** | 810 (51%) | 1336 (48%) | 777 (51%) | 440 (54%) | 466 (53%) | 464 (48%) | 511 (50%) | 460 (50%) | 373 (50%) | 495 (53%) | 444 (49%) | 365 (46%) | 346 (46%) | |
| **Thur.** | 727 (53%) | 1190 (50%) | 737 (52%) | 416 (55%) | 432 (52%) | 438 (47%) | 479 (49%) | 452 (52%) | 371 (50%) | 485 (54%) | 454 (47%) | 391 (45%) | 367 (49%) | |
| **Fri.** | 570 (52%) | 904 (49%) | 602 (48%) | 322 (56%) | 329 (53%) | 358 (51%) | 388 (52%) | 393 (51%) | 326 (51%) | 402 (51%) | 378 (49%) | 318 (45%) | 327 (45%) | |
| **Sat.** | 408 (49%) | 738 (49%) | 514 (48%) | 255 (56%) | 256 (54%) | 230 (53%) | 282 (52%) | 293 (53%) | 264 (52%) | 366 (53%) | 375 (49%) | 331 (48%) | 325 (47%) | |
| **Total** | 4655 (51%) | 7685 (49%) | 4730 (50%) | 2596 (56%) | 2689 (53%) | 2728 (49%) | 2953 (50%) | 2835 (51%) | 2356 (50%) | 3099 (53%) | 2965 (49%) | 2492 (46%) | 2449 (48%) | **44232 (50%)** |
| **Subjects** | 1268 (51%) | 2187 (49%) | 1415 (51%) | 726 (57%) | 693 (54%) | 665 (50%) | 723 (51%) | 652 (51%) | 539 (50%) | 677 (54%) | 623 (49%) | 506 (46%) | 506 (47%) | **11180 (51%)** |